# Objective Point Symmetry Classifications/Quantifications of an Electron Diffraction Spot Pattern with Pseudo-Hexagonal Lattice Metric

Peter Moeck[1*] and Lukas von Koch[1,2]

[1] Department of Physics, Portland State University, Portland, Oregon, USA, * pmoeck@pdx.edu
[2] Westside Christian High School, Portland, Oregon, USA

   The recently developed information-theoretic approach to crystallographic symmetry classifications and quantifications [1-4] in two dimensions (2D) from digital transmission electron and scanning probe microscope images is adapted here for the analysis of an experimental electron diffraction spot pattern, Fig. 1, for the first time. Digital input data are in [1-3] considered to consist of the pixel-wise sums of approximately Gaussian distributed noise and an unknown underlying signal that is strictly 2D periodic. Structural defects within the crystals or on the crystal surfaces, instrumental image recording noise, slight deviations from zero-crystal-tilt conditions in transmission electron microscopy (TEM), inhomogeneous staining in structural biology studies of intrinsic membrane protein complexes in lipid bilayers, …, and small inaccuracies in the algorithmic processing of the digital data all contribute to a single generalized noise term. The plane symmetry group and projected Laue class [1,3] (or the 2D Bravais lattice type [4]) that are/(is) "anchored" [1] to the least broken symmetries are identified as genuine in the presence of generalized noise. More severely broken symmetries that are not anchored to the least broken symmetries are identified as pseudosymmetries.

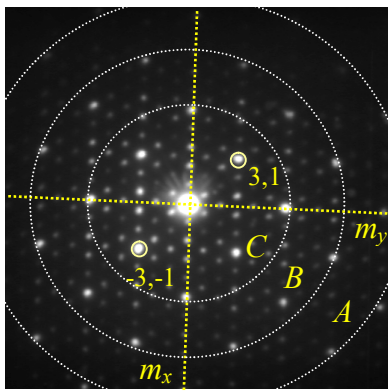

**Figure 1.** Experimental electron diffraction spot pattern from [5]. The quasi-horizontal mirror line is $m_y$ (..m) and its quasi-vertical counterpart is $m_x$ (.m.). Very low intensity spots are not readily visible as this is an 8-bit dynamical range pattern. The added circles represent the concentric circular selection regions as defined in Table I. (Their locations are only approimate.) A Friedel pair of spots is indexed according to the primitive (pseudo-hexagonal) lattice parameters given below.

   The electron diffraction pattern in Fig. 1 is from a $Ba_3Nb_{16}O_{23}$ crystal, space group *Cmmm*, Z = 2. In the crystallographically exact [001] zone axis orientation, i.e. at zero-crystal-tilt, an experimental transmission electron diffraction pattern from a plane-parallel slab of a highest crystalline quality $Ba_3Nb_{16}O_{23}$ crystal would (in an ideal TEM) feature point symmetry group *2mm*. An atomic resolution TEM image of such a crystal in that precise orientation would feature the site symmetries *2mm*, *2*, *.m.* ($m_x$), *..m* ($m_y$), and *1* at the prescribed locations (Wyckoff positions [6]) in a rectangular-centered unit cell that features plane symmetry group *c2mm*. Crystals from which transmission electron diffraction patterns were recorded are, however, hardly ever oriented exactly along low indexed zone axes. They also typically feature shapes other than that of plane-parallel plates and contain structural defects. As a result, the point symmetries in electron diffraction spot pattern and non-overlapping featureless (blank) disk patterns are often lower than what is predicted for orthogonal projections along low indexed zone axes on theoretical grounds.
   In a well known electron crystallography monograph [5], the point symmetry of the electron diffraction pattern in Fig. 1 was "declared" to be *2mm*, although the breakings of two approximately orthogonal mirror lines and a two-fold rotation point (that is slightly off center) are clearly visible. In the context of the original crystal structure determination of $Ba_3Nb_{16}O_{23}$ based on complementing neutron and X-ray diffraction data [7], an "approximate" *2mm* point symmetry classification is, however, justified. While typical for the state-of-the-art in the field, such an approximate point symmetry classification lacks (*i*) any measure of uncertainty (e.g. classification confidence level) and (*ii*) a quantification of the probability of this particular symmetry classification being the best (in some quantifiable, e.g. information-theoretic, sense) with respect to its alternatives. Such alternatives are, on the one hand, point symmetries *2*, *.m.*, and *..m* as maximal subgroups of *2mm*. On the other hand, *6mm* is a minimal supergroup of *2mm* in three symmetry equivalent settings, just as *6* is with respect to *2*. Note that the direct space lattice parameters of the crystal that underlies the diffraction pattern in Fig. 1 are $a = 12.46 \pm 0.2$ Å, $b = 12.41 \pm 0.2$ Å, and $\gamma = 119.5 \pm 1.0°$, as extracted with the well known electron crystallography program CRISP/ELD [5] in its default setting. The existence of both a hexagonal unit cell and merohedry [6] at the hexagonal point symmetry group level can, therefore, not be ruled out on experimental grounds.

Objective crystallographic symmetry classifications/quantifications have been undertaken for larger and smaller circular regions of the diffraction pattern in Fig. 1. The CRISP/ELD 2.1 program [5] was used in the "shape integration" mode to extract the intensities of the electron diffraction spots. These intensities were exported as ∗.hke files. Two of these files were manually amended for *2mm*-symmetry-equivalent missing spots (with intensity zero) and two of them were restricted in their Abbe resolution. Details of these restrictions and their consequences are given in Table I.

| ∗.hke file restriction details | Region A | Region B | Region C |
|---|---|---|---|
| Minimal d-spacing in Å (nominal Abbe resolution) | 0.85 | 1.25 | 2.03 |
| Spot intensity at minimal d-spacing (based on 256 gray levels) | 5.2 | 6.4 | 23.8 |
| Total # of spots | 436 | 256 | 108 |
| Laue indices for minimal d-spacing | $(11,3)_{primitive}$ and $(11,17)_{centered}$ | $(5,5)_{primitive}$ and $(5,15)_{centered}$ | $(4,-6)_{primitive}$ and $(4,-8)_{centered}$ |

**Table I:** Details on the selected concentric circular regions in Fig. 1.

The numerical results of our point symmetry classifications/quantifications in the default primitive (pseudo-hexagonal) indexing of the CRISP/ELD program are given in Tables II to IV for the three regions that are marked in Fig. 1. They were calculated from the ∗.hke files by programs that the second author of this paper wrote.

| Point group | Normalized SSR | G-AIC values | Geometric Akaike weights (%) | Classical $R_{sym}$ (%) |
|---|---|---|---|---|
| *2* | 1.18815117 | 2.277461629 | 20.37217561 | 19.2 |
| *.m.* | 0.544655229 | **1.633965687** | **28.10417102** | 13.2 |
| *..m* | 0.7914292076 | 1.880739666 | 24.84188179 | 14.9 |
| *2mm* | 1.261740485 | 1.806395714 | 25.78268098 | 20.0 |
| *6* | 9.633251966 | 9.996355452 | 0.4294384378 | 56.3 |
| *6mm* | 9.635775834 | 9.817327577 | 0.4696521593 | 56.3 |

**Table II:** Results for region A, 436 spots, K-L best point group *..m* in bold font, primitive indexing.

| Point group | Normalized SSR | G-AIC values | Geometric Akaike weights (%) | Classical $R_{sym}$ (%) |
|---|---|---|---|---|
| *2* | 0.8774512281 | 2.133316644 | 20.72312299 | 15.1 |
| *.m.* | 0.5001790517 | 1.756044468 | 25.02527259 | 11.7 |
| *..m* | 0.5093608102 | 1.765226226 | 24.9106479 | 11.3 |
| *2mm* | 0.9418990621 | **1.56983177** | **27.46720003** | 15.9 |
| *6* | 8.013230141 | 8.431851947 | 0.8886804812 | 52.3 |
| *6mm* | 8.016578923 | 8.225889826 | 0.9850760216 | 52.3 |

**Table III:** Results for region B, 256 spots, K-L best point group *2mm* in bold font, primitive indexing.

| Point | Normalized SSR | G-AIC values | Geometric Akaike weights (%) | Classical $R_{sym}$ (%) |
|---|---|---|---|---|
| *2* | 0.4362919888 | 0.847183493 | 22.21634502 | 13.6 |
| *.m.* | 0.3245002274 | 0.7353917316 | 23.49350878 | 11.1 |
| *..m* | 0.2054457521 | **0.6163372563** | **24.93447537** | 8.5 |
| *2mm* | 0.4831189841 | 0.6885647362 | 24.05006403 | 14.2 |
| *6* | 4.994388369 | 5.131352204 | 2.608418046 | 46.0 |
| *6mm* | 4.995938079 | 5.064419997 | 2.697188755 | 46.0 |

**Table IV:** Results for region C, 108 spots, K-L best point group *.m.* in bold font, primitive indexing.

With a reduced number of electron diffraction spots, i.e. reduced Abbe resolution, one expects that both the normalized sum of squared residuals (N-SSR) and the classical $R_{sym}$ values acquire lower numbers/percentages, as observed in these three tables. The $R_{sym}$ [5] values identify the least broken point symmetry group (anchoring group) for regions A and C in Fig. 1 correctly, but not the point group that the digital data actually supports best in the information-theoretic (or any other objective, i.e. researcher independent) sense for region B.

The key results of our study (and our scientific progress with respect to relying on the classical $R_{sym}$ values for classifications of electron diffraction spot patterns into point symmetry groups) are given in the third and fourth columns of Tables II to IV. These are the geometric Akaike Information Criterion (G-AIC) values (which are model-selection-bias corrected residuals [1,3]) and the geometric Akaike weights (which are the probabilities that a certain point symmetry group is the Kullback-Leibler (K-L) best representation within a set of alternative point groups [2,3]) for the three circular regions of the electron diffraction pattern in Fig. 1. The geometric Akaike weights in these tables add up to 100% for the whole set of analyzed point symmetry groups (as probabilities for complete sets of alternatives always must). In order to highlight this additive feature, these value were not rounded and are just presented as obtained from the above mentioned computer programs of the second author of this paper. (The $R_{sym}$ values are rounded and do not feature this additive feature as they are only relative symmetry deviation measures of each point symmetry group individually to the experimental data.)

For region B, point symmetry *2mm* is the K-L best group, featuring the highest geometric Akaike weight. The average confidence level [1] for preferring point symmetry *2mm* over its three maximal subgroups is 38.83%. Both of the information-theoretic symmetry deviation quantifiers allow not only for objective point symmetry classifications in the presence of generalized noise but also for their quantifications. Tables II to IV illustrate the high sensitivity of the point group classifications/quantifications to the selected three concentric regions of the pattern in Fig. 1. The classical $R_{sym}$ values vary as well with these regions but do not allow for objective classifications in the first place.

When larger circular regions of an electron diffraction spot pattern are analyzed, one would expect that the K-L best point symmetry group stands out more clearly with respect to its maximal subgroups and/or the pseudosymmetry groups. More data points equate to more information that is to be extracted and quantified in the presence of generalized noise. This connection is reflected in information-theoretic results with high probabilities for symmetry alternatives that are highly likely and low probabilities for symmetry alternatives that are highly unlikely, given the prevailing generalized noise level, as shown in Tables II to IV.

It is clear from our point symmetry analysis that the electron diffraction pattern in Fig. 1 does not feature a six-fold rotation point so that the apparent metric symmetry of the extracted lattice parameters signifies only a strong translational pseudosymmetry [8]. The diffraction pattern needs, accordingly, to be re-indexed for a rectangular-centered Bravais lattice with parameters $a = 12.45 \pm 0.2$ Å, $b = 21.60 \pm 0.2$ Å, and $\gamma = 89.5 \pm 1.0°$ (before symmetrization to 90°), as obtained with CRISP/ELD in good agreement to neutron and X-ray diffraction results [7]. Re-indexed Laue indices for the electron diffraction spots with the minimal d-spacings for the three concentric regions in Fig. 1 are given in the last row of Table I.

The objective point symmetry classification/quantification results are theoretically independent of the labels of the electron diffraction spots, but it is interesting to compare the results in Tables III and V. The latter table is for region B of Fig. 1 (just as Table III), but for a rectangular-centered indexing of the 256 electron diffraction spots within a circular region that corresponds to an Abbe resolution of 1.25 Å.

| Point group | Normalized SSR | G-AIC values | Geometric Akaike weights (%) | Classical $R_{sym}$ (%) |
|---|---|---|---|---|
| *2* | 0.8735826567 | 2.151534678 | 20.85116526 | 15.4 |
| *.m.* | 0.5287317871 | 1.806683809 | 24.77500468 | 11.9 |
| *..m* | 0.5176219373 | 1.795573959 | 24.91301093 | 11.6 |
| *2mm* | 0.9584640163 | **1.597440027** | **27.50745779** | 16.1 |
| *6* | 7.955217256 | 8.381201263 | 0.9254978613 | 52.2 |
| *6mm* | 7.95839763 | 8.171389634 | 1.02786348 | 52.2 |

**Table V:** Additional results for region B, 256 spots, K-L best point group *2mm* in bold font, rectangular-centered indexing.

The CRISP/ELD program extracted slightly different intensities for the same spots when the rectangular-centered indexing that corresponds to the lattice parameters of the last paragraph was used. (Similarly, other electron crystallography programs than CRISP/ELD are bound to give slightly different results as demonstrated for lattice parameter extractions in [8].)

The average confidence level for preferring point symmetry group *2mm* over its three maximal subgroups is 40.08% for the rectangular-centered indexing of region B. There is, thus, a difference of 1.25% to the analysis for the hexagonal indexing for the same region in Fig. 1. Without being privy to the details of the electron diffraction spot-shape integration extraction routine of CRISP/ELD it is not knowable which of the two average confidence levels is more accurate. Remarkably, the geometric Akaike weights for point symmetry group *2mm* in Tables III and V differ by only 0.04%. This weight may, thus, be a more robust quantifier of the point symmetry of region B of the electron diffraction pattern in Fig. 1.

A human symmetry classifier would probably have concluded that for region A of Fig. 1, i.e. the whole pattern, the symmetry around the almost vertical mirror line, $.m.$ ($m_x$), is less broken than the symmetry around the almost horizontal mirror line, $..m$ ($m_y$). This is because there are visibly more electron diffraction spots in the lower half of Fig. 1 than in the upper half.

For region C in Fig. 1, i.e. the innermost part of the electron diffraction pattern, a human classifier would probably have concluded that the symmetry around the almost horizontal mirror line, $..m$ ($m_y$) is less broken than the symmetry around the almost vertical mirror line, $.m.$ ($m_x$). This can be appreciated by the spots 3,-1 and 3,-2 (right-hand side of the quasi-horizontal mirror line) being visibly less intense than the -3,1 and -3,2 spots (left-hand side of that mirror line). Quantitative symmetry classifications are obviously more accurate than what a human classifier may come up by visual inspection alone. For the presented electron diffraction pattern symmetry study, it is kind of reassuring that the newly written computer programs that we used for our point symmetry quantifications are in accord with qualitative-visual symmetry inspections.

The point symmetry classification results of Table II are suggestive of a slight mis-orientation of the crystal away from the exact [001] zone axis. The tilt axis of the sample holder should be orientated more or less horizontally in the diffraction pattern in Fig. 1. The classification results in Tables II to IV are in aggregate indicative of the crystal (area from which the diffraction pattern was taken) featuring a non-trivial real structure that may include merohedral or pseudo-merohedral twinning. Complementing symmetry classification and quantification analyses of the diffraction pattern in Fig. 1 are provided in [9].

Our point symmetry quantification study of an electron diffraction spot pattern is highly topical because a novel contrast mechanism for 2D scanning transmission electron microscopy (STEM) on the nodes of a 2D net, commonly called 4D STEM, was recently demonstrated by other authors [10]. That contrast mechanism relies on crystallographic point/site symmetries being treated as continuous features in direct space. Their method employs, however, a classifier for point symmetries [11] that has no objective way of dealing with the hierarchical aspects [12] of these symmetries, i.e. the well known point symmetry inclusion relationships [6]. Their classifier has also no inbuilt feature to distinguish between genuine symmetries and pseudo-symmetries in experimental data, which is always noisy.

Our point symmetry classification/quantification method overcomes both of these shortcomings at once and is, therefore, poised to make a contribution to the future refinement of the novel [10] imaging mode for 4D STEM imaging with fast pixelated detectors. With an almost parallel scanning nano-beam, one can expect a high sensitivity of the information-theoretic point symmetry quantifiers for different locations within the unit cell of a crystal with sufficiently large unit cell. One could, for example, color code an atomic resolution STEM image of $Ba_3Nb_{16}O_{23}$ in [001] zone axis orientation with the local values or derivatives of the geometric Akaike weights for point group symmetries *2*, *.m.*, *..m*, and *2mm* in order to reveal experimentally quantified crystallographic site symmetries in direct space within the average unit cell and/or each individual unit cell. A high sensitivity to local symmetry variations will make the new STEM contrast mode [10] more useful. The usage of objective symmetry quantifications is bound to become the preeminent condition of the establishment of that novel contrast mode as an industry-wide accepted standard.